\documentclass{article}
\usepackage{geometry}
\usepackage{graphicx} % Required for inserting images 
\usepackage{multicol}
\usepackage{physics}
\usepackage{mathtools}
\usepackage{amsmath}
\usepackage{amssymb}
\usepackage{amsthm}
\usepackage{xcolor}
\usepackage{appendix}
\usepackage{authblk}

\usepackage{hyperref}
\hypersetup{
    colorlinks=true,
    linkcolor=blue,
    filecolor=magenta,      
    urlcolor=cyan,
    pdftitle={Overleaf Example},
    pdfpagemode=FullScreen,
}

\begin{document}

% \title{Gravitational Wave Propagation in a $R^2$-Gravity Condensate of Geometric Origin }
\title{Gravitational Wave Propagation  in a Geometric Condensate  in Starobinsky Cosmology}

%\author{Aurindam Mondal}
%\email{aurindammondal99@gmail.com} 
%\affiliation{Physics and Applied Mathematics Unit, Indian Statistical Institute, 203, Barrackpore Trunk Road, Kolkata, India- 700108.}

%\author{Subir Ghosh}
%\email{subirghosh20@gmail.com}
%\affiliation{Physics and Applied Mathematics Unit, Indian Statistical Institute, 203, Barrackpore Trunk Road, Kolkata, India- 700108.}

\author[1]{Aurindam Mondal\thanks{\texttt{aurindammondal99@gmail.com}}}
\author[1]{Subir Ghosh\thanks{\texttt{subirghosh20@gmail.com}}} 
\affil{\makebox[\textwidth][c]{\small{$^{1}$Physics and Applied Mathematics Unit, Indian Statistical Institute,}} \newline 
          \makebox[\textwidth][c]{\small{203, Barrackpore Trunk Road, Kolkata, India-         700108.}}}
% \makebox[\textwidth][c]{\affil[1]{\small{Physics and Applied Mathematics Unit, Indian Statistical Institute}}} \newline   
% \makebox[\textwidth][c]{\affil[1]{\small{203, Barrackpore Trunk Road, Kolkata, India- 700108.}}}

\date{\vspace{-1.3 cm}}

\maketitle 

\abstract{In this paper we propose a new paradigm for cosmology: a time dependent scalar condensate background originated from the quadratic $(R + \alpha R^2)$ Starobinski model, where $R$ is the Ricci scalar and $\alpha$ the coupling constant. In weak gravity limit the system decouples into a conventional graviton and a higher derivative scalar. It was shown earlier through works from our group \cite{ssg,sg,us} that the latter can sustain an oscillatory lowest energy configuration or a {\it{Geometric Condensate}} as it consists entirely of metric degrees of freedom. 

In the present work, we study Gravitational Wave propagation in this condensate background. We show that the explicit time dependent nature of the condensate can generate spatial curvature and radiation-like contributions in the scale factor evolution in FLRW cosmology. Subsequently the condensate leaves its signature on the Gravitational Wave  profile as it propagates in the condensate modified FLRW spacetime. The wave profile is calculated  analytically in terms of Whittaker functions. The main novelty of the Geometric Condensate scheme is that no external (condensate) matter from outside has been considered.}

%\tableofcontents 

\section{Introduction}

The direct detection of the Gravitational Wave (GW) event GW150914 reported by LIGO \cite{ligo}, and the later by others  \cite{other}, have ushered in a new window to astrophysics and cosmology. It is expected that GW will bear imprints, not only of the source but also of the interstellar medium through which it has to pass in order to reach earth based observatories. Among various observables concerning the GW, its amplitude, frequency and speed are of fundamental importance. In the present work we will study GW propagation through a novel background: a {\it{Higher Derivative Scalar}} (HDS) condensate or {\it{Geometric Condensate}} (GC) proposed in \cite{ssg,sg} (and further developed in \cite{us}, generated from the Starobinski-like quadratic gravity model \cite{star}). Studies of GW in generic $f(R)$ models have appeared in \cite{capo}. Apart from its crucial role in inflationary cosmology, the Starobinski model is one of the very few of so called $f(R)$-gravity models that is free from Ostrogadsky ghosts (see for example \cite{st}), as explained in detail in \cite{alv}. However, the feature of the Starobinski $(R+\alpha R^2)$ model that is relevant to our work, is its linearized version consists of the usual Einstein Gravity with a spin-$2$ particle (graviton) and a decoupled, spin-$0$ {\it{ Higher  Derivative  Scalar}} \cite{alv}. The latter excitation might have been considered as an embarrassment due to its persistence in the low energy regime. In this perspective our interpretation of HDS  giving rise to the condensate \cite{ssg,us} tends to solve the extra degree of freedom issue. We name this as {\it{Geometric Condensate}} (GC). In \cite{ssg} it was established that the higher derivative scalar is capable of generating a lowest energy configuration with a non-trivial (periodic) time dependence \cite{tc} and that this condensate can replace the pure vacuum. In \cite{us} we considered a scenario where the energy-momentum tensor corresponding to this condensate back-reacts  on conventional Einstein Gravity and discussed effects of the condensate on evolution of the scale factor in a Friedmann-Lemaitre-Robertson-Walker (FLRW) framework. In the present work we probe the effects of condensate in another sector: propagation of GW in this GC background. Under certain well-motivated approximations, we solve the linearized GW equation with the GC playing the role of ``matter". \\ 

We emphasize that {\it{no matter contribution is introduced from outside and the GC (acts as a ``matter" distribution) is composed from the quadratic $(R + \alpha R^2)$ term of Starobinski model}} \cite{alv,ssg,us}. Keeping the above in mind, in a nutshell, the interesting results that we have obtained are the following: 
\begin{itemize}
    \item An effective Cosmological Constant-like contribution appears due to the AdS background \cite{alv,ssg}. The new results are the generation of effective spatial curvature-like contribution and an effective radiation like contribution, both appearing  entirely due to the time-dependent GC. As already mentioned, We stress that the non-trivial time dependence of GC is responsible for the above mentioned  physically interesting contributions and no matter component is added from outside.   
    \item The new and most important contribution of the present work is the study of the propagation of Gravitational Waves in this non-trivial background. We analytically compute, under certain well motivated approximation, the GW profile from which effect of the GC can be directly observed. The GW carries the parameter $\alpha$ - signature of the quadratic $R^2$ term - through the GC. The GC creates additional structure over and above  the  attenuated GW, that is due to  cosmological expansion.  
 \end{itemize}  

Indeed, we do not claim that our results will match with realistic observations in cosmology since we have not considered the true matter and radiation sectors but the theoretical outcome from our condensate model is appealing and merits further study. In passing we point out another interesting perspective: the GC has close connection with systems enjoying Time Crystal features \cite{tc}.

Before concluding this section we point out several ideas of introducing condensates of various origin, existing in the literature. However, we stress that the fundamental difference between all these condensate frameworks and our Geometric Condensate approach is that in all the former cases the condensate is constructed out of external matter introduced from outside whereas in the latter formalism the GC is built within the $R^2$-gravity scenario and is composed entirely out of metric degrees of freedom, justifying the name of Geometric Condensate. 

\begin{itemize}
    \item \textit{Bose-Einstein Condensate:} \\ 
In the review \cite{h1}, Bose–Einstein condensate (BEC) formed out of   collective spin-wave excitations or  magnons is considered and its relevance in diverse physical phenomena, from  particle physics and cosmology to vortex dynamics, and most interestingly in Time Crystal-like new phases of condensed matter, are discussed. Closest to our work in spirit is possibly \cite{h2} that proposes a Ghost condensation (in analogy to  the Higgs mechanism) in general relativity that generates a vacuum expectation value of the derivative of a scalar field. Again in \cite{h3}, the relevant BEC is proposed to be induced from fermionic neutralinos that can form  Cooper pairs that behave  collectively as a boson field when the temperature of the Universe reaches a critical value. In another variant, \cite{h4}  considers the possibility of dark matter appearing as  a non-relativistic, Newtonian gravitational BEC condensate, whose density and pressure satisfy a barotropic equation of state. On the other hand, \cite{h5} examines  the possibility of a unified picture of the dark sector of the universe where BEC of light bosons can account for cold dark matter and Dark Energy both emerging from the same source. In \cite{h6} suggests that the axion, (itself a popular cold dark matter candidate), might evolve into a BEC driven by ``gravitational thermalisation" in the radiation dominated era. \cite{h7} assumes that bosons forming Dark Matter can form BEC below a critical temperature and later can coexist with Dark Matter to form a Gross–Pitaevskii–Poisson system. 

    \item \textit{Other condensate models:} \\ 
Baryon asymmetry and other issues are addressed in O'Raifeartaigh models \cite{h9} where  SUSY breaking associated with inflation can turn into a condensate with large expectation value and can  carry significant fractional baryon number that survives till date. Still another model \cite{h10} considers a spatially homogeneous isotropic mode of Yang-Mills field condensate and studies the dynamics in an expanding Universe. 

    \item \textit{Group Field Theory Condensate:} \\ 
Another formalism where condensate plays an important role and has created a lot of interest is Group Field Theory (GFT) condensate cosmology \cite{c1} (for a review see \cite{c2}). The essential idea is to replace the spacetime continuum by a group manifold inhabited by degrees of freedom of discrete and combinatorial nature, endowed with a peculiar type of
(combinatorially) non-local interaction. The continuum geometry is recovered via a phase transition to a condensate phase. One of the successes of this framework is the reappearance of Friedmann-like dynamics of an emergent homogeneous and isotropic geometry.
\item \textit{Ghost Condensate:} \\
In another variant of condensate scenario, it has been proposed \cite{ghost} that, scalar field models having  higher-derivative kinetic terms, can  couple to gravity to induce a  vacuum that, is   ghost-free but has features similar to a Cosmological Constant. It is to be noted that the  null energy condition of general relativity is violated in these models.
\end{itemize}
From the above discussion it is clear that there are diverse condensate models in cosmological context, each having some positive points but a common weakness of all of them is that the generation of the condensates themselves rely on  some exotic mechanisms or unsubstantiated matter content. The present GC scores above the others precisely in this context: we only introduce the quadratic ($R^2$) Starobinski extension on top of  Einstein gravity and that is sufficient to induce the GC in purely conventional physics principles.    

The paper is organized as follows:  In Section 2 we quickly recall the generation of the Geometric Condensate (GC) in Starobinski like quadratic $R^2$-cosmology. In Section 3 we analyse gravitational wave propagation in a generic curved background. Section 4 comprises of detailed study of scale factor and Hubble parameter in presence of GC background. Section 5 provides explicit analysis of gravitational wave propagation in the GC background. Section 6 is devoted to discussions of the results, in particular the effects of the condensate in gravitational wave properties. We also point out several open questions.

\section{A quick recap of the Quadratic Gravity and ensuing Geometric Condensate}

The specific form of Quadratic Gravity we consider is
 given by \cite{alv}
\begin{equation}
A=\frac{c^4}{16\pi G}\int d^4x~ \sqrt{-g} \hspace{0.1cm} \big[R+ \alpha R^2-2\Lambda \big]
\label{4}
\end{equation}
with positive $\alpha$  to avoid tachyonic excitations.  We consider the unit system $l~(\mbox{length})$, $m~(\mbox{mass})$, $t~(\mbox{time})$ so that $[\alpha]=[l^{2}]$, $[R]=[l^{-2}]$ and the scale factor $a(t)$ is dimensionless which will appear in the study of GW propagation through FLRW spacetime. 

In order to find the scalar tensor decomposition of the above action, one has to introduce linear order perturbation of the metric tensor as follows.
\begin{equation}
    g_{\mu\nu} \hspace{0.1cm} = \hspace{0.1cm} \widetilde{g}_{\mu\nu} + h_{\mu\nu} 
\end{equation}
Here $\widetilde{g}_{\mu\nu}$ be the arbitrary background metric (although we restrict it to dS or AdS) and $h_{\mu\nu}$ be the small perturbations on top of this smooth background. The perturbed metric can be further decomposed in terms of its irreducible components. 
\begin{equation}
    h_{\mu\nu} \hspace{0.1cm} = \hspace{0.1cm} h^{TT}_{\mu\nu} + \big(\widetilde{\nabla}_{\mu}v_{\nu} + \widetilde{\nabla}_{\nu}v_{\mu} \big) + \big(\widetilde{\nabla}_{\mu} \widetilde{\nabla}_{\nu} - \frac{1}{4} g_{\mu\nu} \widetilde{\Box} \big) v + \frac{1}{4} \widetilde{g}_{\mu\nu} h 
\end{equation}
Here $h^{TT}_{\mu\nu}$ represents the traceless-transverse component of the perturbed metric satisfying the condition $\widetilde{\nabla}_{\mu} h^{\mu\nu}_{TT} = 0$ \hspace{0.1cm} and \hspace{0.1cm} $\widetilde{g}^{\mu\nu} h^{TT}_{\mu\nu} = 0$, $v_{\mu}$ represents the vector part of the perturbed metric satisfying the constraint $\widetilde{\nabla}_{\mu} v^{\mu} = 0$ and $v$ is the scalar degree of freedom of the same. 
In \cite{alv}, the authors proved that this $(R + \alpha R^2)$ action decouples to conventional gravity theory (with a spin $2$ graviton) along with a HDS (modulo surface terms) up to second order perturbative expansion of the action. Incidentally, this is a manifestation of the observation that any generic $f(R)$ gravity can always be expressed as a scalar-tensor theory. The final decoupled form of the action is expressed as, 
\begin{eqnarray}
    A &=& \frac{c^{4}}{16\pi G} \int d^{4}x \sqrt{-\widetilde{g}} \hspace{0.1cm} \bigg[\frac{1}{4} \big(1 + 32\pi G\alpha\widetilde{R}_{0} \big) \hspace{0.06cm} h^{\mu\nu}_{TT} \bigg(\widetilde\Box - \frac{\widetilde{R}_{0}}{6} \bigg) h^{TT}_{\mu\nu} \nonumber \\ 
    && + \hspace{0.1cm} 9\pi G\alpha \bigg(\widetilde{\Box}\phi + \frac{\widetilde{R}_{0}}{3} \phi \bigg)^{2} - \frac{3}{32} \big(1 + 32\pi G\alpha\widetilde{R}_{0} \big) \hspace{0.06cm} \phi \bigg(\widetilde{\Box} + \frac{\widetilde{R}}{3} \bigg) \phi \bigg]
\end{eqnarray}
where $\widetilde{\Box}$ is the D'Alembertian operator defined with respect to the background metric. We consider $\widetilde{R}_{0}$ to be the constant curvature corresponding to the maximally symmetric background spacetime $\widetilde {g}_{\mu\nu}$ \cite{alv}.
In terms of irreducible metric components, the higher derivative scalar $\phi$ is parametrised as $\phi := \big(h - \widetilde{\Box}v \big)$. After some manipulations, the scalar part of the above action is expressed as, (equation (3.51) of \cite{alv}, see also \cite{ssg} for further details) ,
\begin{eqnarray}
A_{GC}= \bigg(\frac{9 c^4}{128 \pi G} \bigg)\frac{1}{2}\int d^4x~\sqrt{-\widetilde g} \Bigg[\alpha \bigg\{(\widetilde{\Box} \phi )^2  +\frac{\widetilde{R}_{0}}{3}\phi\widetilde{\Box}\phi \bigg\} - \frac{1}{6} \bigg(\phi\widetilde{\Box}\phi + \frac{\widetilde{R}_{0}}{3} \phi^2 \bigg) \Bigg] .
\label{8aa}
\end{eqnarray}
After dropping off  a total derivative term, the above action (\ref{8aa}) coincides with the generic action of an arbitrary higher derivative scalar, considered in the reference \cite{Gibbons}. This generic action is written in the following manner. 
\begin{eqnarray}
    S &=& -\frac{1}{2} \int d^4x~\sqrt{-g} \bigg[(\Box\phi)^2+(m_1^2+m_2^2)g^{\mu\nu} \nabla_{\mu}\phi \nabla_{\nu}\phi + m_1^2m_2^2 \hspace{0.06cm} \phi^2 \bigg]
\label{1}
\end{eqnarray} 
Here $\{m_{1},m_{2} \}$ are two distinct, real quantities having  mass dimension. For this above generic action (\ref{1}), the usual procedure of computing energy-momentum tensor yields the following result, 
\begin{eqnarray}
    T_{\mu\nu} &=& \bigg[ -(\partial_\mu\Box\phi)\partial_\nu\phi -(\partial_\nu\Box\phi)\partial_\mu\phi + \frac{g_{\mu\nu} (\Box\phi)^2}{2} -\frac{m_1^2m_2^2}{2} g_{\mu\nu}\phi^2 \nonumber \\ 
    && \hspace{0.1cm} + \hspace{0.1cm} g_{\mu\nu} g^{\alpha\beta}(\partial_\alpha\Box\phi )\partial_\beta\phi + \big(m_1^2+m_2^2 \big) \bigg(\partial_\mu\phi\partial_\nu\phi -\frac{1}{2}g_{\mu\nu}g^{\alpha\beta}\partial_\alpha \phi\partial_\beta\phi \bigg) \bigg] .
\label{10aa}
\end{eqnarray}
For our model, the appropriate expressions of $\{m_{1},m_{2} \}$ can be found in a straightforward way by mapping  the action of the Geometric Condensate given in the Eq. (\ref{8aa}), to the generic higher derivative scalar action (\ref{1}),  introduced in \cite{Gibbons}. These expressions are the following 
\begin{eqnarray}
    (m_1^2+m_2^2)=\bigg(\frac{1}{6\alpha}-\frac{\widetilde{R}_{0}}{3} \bigg) \hspace{0.4cm}; \quad 
    m_1^2m_2^2=\bigg(-\frac{\widetilde{R}_{0}}{18\alpha} \bigg) .
    \label{par}
\end{eqnarray}
After putting these above expression of the mass terms in the Eq. (\ref{10aa}), one attains the following form of energy-momentum tensor for our given GC. 
For simplification, we are going to remove the tilde symbol from the quantities depend on background metric $\widetilde{g}_{\mu\nu}$. 
%\begin{widetext}
\begin{eqnarray}
T_{\mu\nu} &=& \bigg(\frac{9 c^4}{128 \pi G} \bigg) \bigg[\alpha \bigg\{ (\partial_\mu\Box\phi)\partial_\nu\phi + (\partial_\nu\Box\phi)\partial_\mu\phi - \frac{1}{2}g_{\mu\nu}(\Box\phi)^2 - g_{\mu\nu} g^{\alpha\beta}(\partial_\alpha\Box\phi) \partial_\beta\phi \bigg\}  \nonumber \\ 
&& - \hspace{0.1cm} \frac{\widetilde{R}_{0}}{36} \hspace{0.1cm} g_{\mu\nu}\phi^2 + \left(\frac{\alpha \widetilde{R}_{0}}{3} - \frac{1}{6} \right) \left(\partial_\mu\phi\partial_\nu\phi -\frac{1}{2} g_{\mu\nu} g^{\alpha\beta} \partial_\alpha \phi\partial_\beta\phi \right) \bigg]
\label{10a}
\end{eqnarray}
%\end{widetext}

that satisfy the (on-shell) conservation law $\nabla _\mu T^{\mu\nu}=0$.
The dimension of $[T_{\mu\nu}] = (ml^2/t^2) (1/l^3) = \mbox{Energy~density}.$ Note that for $\alpha =0$, the $T_{\mu\nu}$ reduces to that of the conventional scalar field.

\textbf{Birth of the Geometric condensate}: The energy density $T_{00}$ is minimized in momentum space to obtain the Fourier mode frequencies corresponding to the geometric condensate (for details see \cite{ssg,us}).
\begin{equation}
\phi = \cos{(\omega\eta)} \hspace{0.3cm} ; \hspace{0.3cm}  \omega=ca\sqrt{\frac{\widetilde{R}_{0}}{3}}
    \label{cond}
\end{equation}
with $\eta$ denoting conformal time. During minimization, we had used the FLRW spacetime as the background metric over which the geometric condensate is distributed thoroughly. Note that the frequency is not  constant as it depends on the scale factor $a(\eta)$.  This is the all important geometric condensate whose possible existence was established earlier \cite{sg,ssg,us}. In a previous work \cite{us} we had studied cosmological evolution in the presence of the condensate background. In the present work we deal with the effects of the geometric condensate on GW.

\section{Gravitational waves on a curved background}

In order to study the effect of GC (or any generic matter distribution) a straightforward procedure would be to consider the inhomogeneous linearized wave equations corresponding to a first order perturbation of the metric, with the energy-momentum tensor pertaining to the matter,  as the source. However, this can be awkward in our case where the GC is not localized in a finite volume but pervades the spacetime and hence Green's function techniques will be difficult to exploit. So we  follow another  approach used  in \cite{nb}. In this two step procedure, one first solves the background Einstein field equation with the GC. In  case of a short wavelength or high frequency approximation, (where the GW wavelength is small compared to the typical length scale of the background), as assumed here, the GC acts as the source of background Einstein field equation, leaving the perturbative Einstein equation of the first order, i.e. the GW equation, source-free. This constitutes the first step. Effects of the GC appear in the (generically spacetime dependent) coefficients of the otherwise free GW equation. Solving the latter is the second step. Some steps are provided in the Appendix A with more details in \cite{nb}.

Final form of the GW equation propagating on a curved background, is given by
\begin{equation}
    \big(\Box\tilde{h}_{\alpha\beta} - 2\overline{R}_{\nu\beta\alpha\mu} \tilde{h}^{\nu\mu} - \overline{R}_{\nu\alpha} \tilde{h}^{\nu}_{\beta} - \overline{R}_{\nu\beta} \tilde{h}^{\nu}_{\alpha} \big) = 0 
    \label{abc}
\end{equation}
where $\overline{R}_{\alpha\beta\mu\nu}$ is the background Riemann tensor of FLRW spacetime in presence of the GC. For the purpose of simplification, we have used the trace reversed tensor $\Tilde{h}_{\alpha\beta}$ whose definition is this: $\Tilde{h}_{\alpha\beta} = \big(h_{\alpha\beta} - \frac{1}{2} \overline{g}_{\alpha\beta
} h \big)$. We will study  GW   propagation through this FLRW background  modified by the GC. The line element of FLRW spacetime (in the absence of any spatial curvature $k=0$) can be written as 
\begin{equation}
    ds^{2} = \big[-dt^{2} + a^{2}(t) \big(dx^{2} + dy^{2} + dz^{2} \big) \big]
    \label{g1}
\end{equation}
where $\{t,x,y,z \}$ being the cosmic time and Certisian coordinates respectively. The functional form of the scale factor $a(t)$ will be calculated  in the next section, as solution of the Friedmann equation in presence of the GC. The GW equation (\ref{abc}) is reduced further utilising a  synchronous gauge in addition with the Traceless-Transverse gauge or $TT$-gauge. The constraints resulting from  synchronous gauge are $\tilde{h}_{0\mu}=\tilde{h}_{1\mu}=0$ and $\tilde{h}_{22}=-\tilde{h}_{33}$ where $\mu=\{0,1,2,3 \}$.  After imposing these conditions on the trace reversed perturbed metric $\tilde{h}_{\alpha\beta}$, we are left with only two non zero, independent degrees of freedom i.e. $\tilde{h}_{22}$ and $\tilde{h}_{23}$ respectively. This gives rise to the following dynamical equations  for the two above mentioned  independent metric components (with over-dot $\dot{a}$ representing cosmic time derivative),  
\begin{eqnarray}
    \Box \tilde{h}_{22} - \bigg(\frac{2\Ddot{a}}{a} + \frac{6\dot{a}^{2}}{a^{2}} \bigg) \tilde{h}_{22} &=& 0 \hspace{0.06cm} , \nonumber \\ 
    \Box \tilde{h}_{23} - \bigg(\frac{2\Ddot{a}}{a} + \frac{6\dot{a}^{2}}{a^{2}} \bigg) \tilde{h}_{23} &=& 0  \hspace{0.06cm} . 
\end{eqnarray}
Since the above equations are identical, it is enough to solve one of them. For convenience we choose an ansatz that the perturbation components are functions of $t, x$ only,  $\tilde{h}_{22}(x^{\mu})=h(t,x)$, leading to the  second order differential equation. 
\begin{equation}
    \bigg(\frac{\partial^{2}h}{\partial t^{2}} \bigg) + \frac{3\dot{a}}{a} \bigg(\frac{\partial h}{\partial t} \bigg) - \frac{1}{a^{2}} \bigg(\frac{\partial^{2}h}{\partial x^{2}} \bigg) - \bigg(\frac{2\Ddot{a}}{a} + \frac{6\dot{a}^{2}}{a^{2}} \bigg) h = 0\label{g2} 
\end{equation}
 Imposing the separation of variable condition $h(t,x)=u(x) v(t)$, (\ref{g2}) reduces to
\begin{eqnarray}
    \bigg(\frac{d^{2}v}{dt^{2}} \bigg) + \frac{3\dot{a}}{a} \bigg(\frac{dv}{dt} \bigg) - \bigg(\frac{2\ddot{a}}{a} + \frac{6\dot{a}^{2}}{a^{2}} - \frac{\Gamma}{a^{2}} \bigg) v &=& 0 \nonumber \\ 
    \bigg(\frac{d^{2} u}{dx^{2}} \bigg) + \Gamma u &=& 0 
    \label{uv}
\end{eqnarray}
where $\Gamma$ is the separation constant   which may take arbitrary values. Clearly $u(x)$ is given by the solution of simple harmonic oscillator.  Remaining task is to solve damped oscillator equation for $v(t)$ with time dependent damping factor and frequency where explicit form of $a(t)$ has to be computed from the background Friedmann equations. Remember one has to consider the existence of GC over the entire background spacetime while solving the Friedmann equations. This will be the first step in the two step procedure.

\section{FLRW cosmology in presence of Geometric Condensate}

Let us now proceed to the first step. In the presence of this above mentioned stress-energy tensor (originated from GC), we are going to analytically solve the Friedmann equations (in the conformal time $\eta$) with the help of some approximations, an obvious one being small $\alpha$, the coupling constant. 
In conformal time, the Friedmann equations in presence of the GC are expressed as follows. 
\begin{eqnarray}
    \left(\frac{a'}{a} \right)^{2} &=& -\frac{8\pi G a^2}{3} \hspace{0.06cm} T^0_0  \hspace{0.1cm} , \label{a4} \\
    \frac{a''}{a}- \left(\frac{a'}{a} \right)^2 &=& \frac{4\pi Ga^2}{3} \big(T^0_0-3T^i_i \big) . \label{a44}
\end{eqnarray} 
After eliminating the second order derivative of the scale factor from the above two equations (\ref{a4}) and (\ref{a44}), one obtains the following quadratic equation of conformal time Hubble parameter $\mathcal{H} = \big(a'/a \big)$. Using explicit forms of $T^{0}_{0}, T^{i}_{i} $ from (\ref{10a}), we get
\begin{eqnarray}\label{Quadratic}
    \left(\frac{a'}{a} \right)^{2} - \frac{3\alpha\phi'\phi''}{4c^2a^6} \left(\frac{a'}{a} \right) - \zeta(\alpha) = 0 
\end{eqnarray} 
where the quantity $\zeta(\alpha)$ is given by
\begin{eqnarray} 
    \zeta(\alpha) &=& \bigg(\frac{R_{0} c^2a^2\phi^2}{192} - \frac{\phi'^2}{64a^2} \bigg) \hspace{0.06cm} \Bigg[1- \frac{3\alpha}{16c^2a^6}\; \left( \frac{\frac{13\phi'^4}{128} -  \frac{R_{0}c^2a^4 \phi'^{2}}{2} - \frac{7}{2} \bigg(\frac{R_{0}c^2a^2\phi^2\phi'^2}{192} \bigg)-\bigg(\frac{\phi''^{2}-2\phi'\phi'''}{2} \bigg)}{\bigg(\frac{R_{0}c^2a^2\phi^2}{192} - \frac{\phi'^2}{64a^2} \bigg)} \right) \Bigg] . \label{p3} \hspace{0.7cm} 
\end{eqnarray} 
Then solving the quadratic equation (\ref{Quadratic}) and restricting to O$(\alpha)$ effects only, we get the following expression of conformal time Hubble parameter \cite{us} ,
\begin{equation}
    \frac{a'}{a} = \Bigg[\frac{3\alpha}{8c^2a^6}\phi'\phi''\pm \sqrt{\zeta(\alpha)} \Bigg] 
    \label{p1}
\end{equation} 
where $\zeta(\alpha)$ is given above in (\ref{p3}). Up to now, we have not used any particular solution of the geometric condensate $\phi$. Let us use the particular geometric condensate solution mentioned earlier in (\ref{cond}) as $\phi=\cos{\big(\omega\eta \big)}$ \cite{us,ssg} in the above equations (\ref{p1}), (\ref{p3}). The time dependent frequency of this solution is $\omega=ca\sqrt{\frac{R_{0}}{3}}$. In order to get a simplified form of the Hubble parameter given in (\ref{p1}), we restrict to a small $\eta$ approximation $\sim$ $\sin{(\omega\eta)} \approx \omega\eta$ and $\cos{(\omega\eta)} \approx 1$, thereby  neglecting   $O\big(\omega^{2}\eta^{2} \big)\approx 0$. However, we emphasize that   $\phi$ still has an explicit time dependence that will prove to be crucial in subsequent analysis and the final outcome. {\footnote{ Note that this is significantly improved from an earlier approximation used by some of us in \cite{us} where we used $\sin{(\omega\eta)} \approx 0$ and $\cos{(\omega\eta)} \approx 1$. On the other hand, in the present, case $\phi$ possess an explicit time dependence.}}

After substituting the approximate solution of the scalar field in the expression of Hubble parameter, one gets the following equation. 
\begin{eqnarray}\label{Explicit_time}
    \frac{a'}{a} &=& \bigg[\bigg(\frac{\alpha R^{2}_{0}c^{2}}{24} \bigg) \frac{\eta}{a^{2}} \hspace{0.04cm} \pm \hspace{0.04cm} \sqrt{\frac{{R_{0}c^{2}}}{192} } \hspace{0.03cm} \hspace{0.03cm} \bigg(a + \frac{\hspace{0.01cm} \alpha R_{0}}{a^{3}} \bigg) \bigg] .
    \label{eta}
\end{eqnarray}
Note the explicit presence of conformal time  $\eta$ in the RHS.
We need to convert this expression from conformal time $\eta$ to cosmic time $t$ since the GW equations (\ref{uv}) are in cosmic time. Indeed, it is also customary to express final results in terms of cosmic time, considered as the physical time. However, the explicit presence of $\eta$ in (\ref{eta}) makes the conversion tricky.  In order to solve this 1st order, non-linear ordinary differential equation, we have to express the conformal time $\eta(t)$ as a function of the scale factor $a(t)$ in this above equation. But it is impossible to establish such a relation between conformal time and scale factor without solving this above equation. For this purpose, let us use the relation: $\eta(t) \propto a(t)$ that is applicable for a radiation dominated universe along with a Cosmological constant term (with appropriate approximation). The exact relation can be established from the definition of conformal time $ \eta(t) = \int ^t \frac{dt}{a(t)} $ which in presence of Cosmological constant and radiation, is given by 
\begin{eqnarray}
    \eta(t) &=& \frac{1}{(2\alpha R_{0})^{1/4}} \int^{t} \frac{dt}{\sqrt{\sinh{\bigg\{\sinh^{-1}{\big(a^{2}_{0}/\sqrt{2\alpha R_{0}} \big)} + \sqrt{R_{0}c^{2}/48} \hspace{0.05cm} (t-t_{0}) \bigg\}}}}.  
\end{eqnarray}
This being an Elliptic integral, the resulting expression is approximated for small $t$ to read 
\begin{eqnarray}\label{conformal_time}
    \eta(t) &\approx& \frac{8\sqrt{6}}{\sqrt{R_{0}c^{2}} \hspace{0.1cm} (2\alpha R_{0})^{1/4}} \hspace{0.1cm} a(t) \hspace{0.1cm} . 
\end{eqnarray}
Hence the replacement of conformal time by this above relation (\ref{conformal_time}), the Hubble parameter in cosmic time reads 
\begin{eqnarray}
    \frac{\dot{a}}{a} &=& \sqrt{\frac{R_{0}c^{2}}{192} + \frac{\alpha R^{2}_{0}c^{2}}{96 \hspace{0.07cm} a^{4}} + \frac{R_{0}c^{2}}{12 a^{2}} \big(\alpha R_{0} \big)^{3/4} } .
    \label{x}
\end{eqnarray}
It has to be kept in mind that (\ref{x}) is not the full story:  in a realistic model of the universe, contributions of matter and (photonic) radiation are indeed present. Hence, as a lowest order correction, contributions of the latter can be simply added inside the square root on the RHS. Further interactions between matter and the GC are assumed to be small and are beyond the scope of the present work.

It is straightforward to solve the differential equation to get the exact form of scale factor.
\begin{eqnarray}
    a(t) &=& \bigg[\big\{a^{2}(t_{0}) + 8 \big(\alpha R_{0} \big)^{3/4} \big\} \cosh{\bigg\{ 2 \sqrt{\frac{R_{0}c^{2}}{192}} (t-t_{0}) \bigg\}} - 8 \big(\alpha R_{0} \big)^{3/4}  \nonumber \\ 
    && + ~\sqrt{a^{4}(t_{0}) + 16a^{2}(t_{0}) \big(\alpha R_{0} \big)^{3/4} + 2\alpha R_{0}} \hspace{0.1cm} \sinh{\bigg\{2 \sqrt{\frac{R_{0}c^{2}}{192}} (t-t_{0}) \bigg\}} \bigg]^{\boldsymbol{1/2}} .
\end{eqnarray}

Here $a(t_{0})$ represents the present day scale factor of the universe. Hubble parameter (or the scale factor $a(t)$) written in cosmic time, directly shows the presence of coupling constant $\alpha$ in the coefficients of $(1/a^{2})$ and $(1/a^{4})$ terms, demonstrating the importance of GC, induced by the $R^{2}$  modification of Einstein-Hilbert action. The  $(1/a^{4})$-term corresponds to an effective radiation-like behaviour, whereas the $(1/a^{2})$-term denotes an effective spatial curvature-like behaviour. The main take home message of this mechanism is that one does not require to put any radiation or spatial curvature like term in the FLRW equations from outside by hand. Signature of explicit time dependent term present in the Hubble parameter written in conformal time, can be found in the coefficient of $(1/a^{2})$ term in the Eq. (\ref{x}). At the present day ($t=t_{0}$), the expression of Hubble parameter can be written as
\begin{eqnarray}
    {H}_0^2 &=& \bigg[\frac{R_{0}c^{2}}{192} + \frac{\alpha R^{2}_{0}c^{2}}{96 \hspace{0.07 cm} {a}_0^{4}} + \frac{R_{0}c^{2}}{12 {a}_0^{2}} \big(\alpha R_{0} \big)^{3/4} \bigg] .
    \label{x1}
\end{eqnarray}

 Furthermore, a restricted form of $a(t)$ is considered for small $t$, that is, for small arguments of the hyperbolic function (up to linear order); $\sinh{(2\sqrt{D}t )} \approx 2\sqrt{D}t,~\cosh{(2\sqrt{D}t)} \approx 1$, with $D=R_0c^2/192$. This yields a scale factor that is a parabolic function of time, $a(t) \approx \sqrt{B + \gamma_{0}t}$ (see Appendix C for computational steps), with the constants defined as 
\begin{eqnarray}\label{coefficients}
    B &=& \bigg[a^{2}(t_{0}) - 2t_{0} \bigg(\frac{R_{0}c^{2}}{192} \bigg)^{1/2} \sqrt{a^{4}_{0} + 16 a^{2}_{0} \big(\alpha R_{0} \big)^{3/4} + 2\alpha R_{0}} \bigg] \hspace{0.1cm} = \hspace{0.1cm} a^{2}(t_{0}) \big(1 - 2H_{0}t_{0} \big) \nonumber \\ 
    \gamma_{0} &=& 2 \bigg(\frac{R_{0}c^{2}}{192} \bigg)^{1/2} \sqrt{a^{4}_{0} + 16 a^{2}_{0} \big(\alpha R_{0} \big)^{3/4} + 2\alpha R_{0}} \hspace{2.0cm} = \hspace{0.2cm} 2 H_{0} \hspace{0.07cm} a^{2}(t_{0})
    \label{yy}
\end{eqnarray}  

It is more convenient to express the Hubble parameter in terms of three density parameters of our cosmological model, $\{\Omega_{\lambda}, \Omega_{R},\Omega_{k} \}$ respectively.  
\begin{eqnarray}
    H(t) &=& \hspace{0.1cm} \sqrt{\frac{R_{0}c^{2}}{192} + \frac{R_{0}c^{2} \big(\alpha R_{0} \big)^{3/4}}{12 \hspace{0.1cm} a^{2}_{0}} \hspace{0.06cm} \frac{1}{(a/a_{0})^{2}} + \frac{\alpha R^{2}_{0} \hspace{0.05cm} c^{2}}{96 \hspace{0.1cm} a^{4}_{0}} \hspace{0.06cm} \frac{1}{(a/a_{0})^{4}} } \nonumber \\ 
    &=& H_{0} \sqrt{\Omega_{\lambda} + \frac{\Omega_{k}}{1+2H_{0}(t-t_{0})} + \frac{\Omega_{R}}{(1+2H_{0}(t-t_{0}))^{2}} } \label{Hubble}
\end{eqnarray}
The above mentioned density parameters are defined in the following manner,
\begin{eqnarray}
    \Omega_{\lambda} = \hspace{0.1cm} \bigg(\frac{R_{0}c^{2}}{192 H^{2}_{0}} \bigg) \hspace{0.3cm} ; \hspace{0.3cm} \Omega_{k} = \hspace{0.1cm} \frac{R_{0}c^{2} \big(\alpha R_{0} \big)^{3/4}}{12 \hspace{0.1cm} a^{2}_{0}H^{2}_{0}} \hspace{0.3cm} ; \hspace{0.3cm} \Omega_{R} = \hspace{0.1cm} \bigg(\frac{\alpha c^{2}R^{2}_{0}}{96 \hspace{0.1cm} a^{4}_{0}H^{2}_{0}} \bigg),
\label{y}
\end{eqnarray}
such that they satisfy with the following constraint
\begin{equation}
\Omega_{\lambda} + \Omega_{k} + \Omega_{R}  = 1. \label{om}
\end{equation}
Here $\Omega_{\lambda}$ corresponds to the fractional energy density of an effective cosmological constant $\lambda = \big(R_{0}c^{2}/34 \big)$,  $\Omega_{k}$ corresponds to an effective spatial curvature $k$  arising from the explicit time dependent portion of the Hubble parameter and finally $\Omega_{R}$ corresponds to an effective radiation-like contribution. 
Below we give two plots of the Hubble parameter as a function of time $``t/t_{0}"$ for different values of density parameters $\Omega_{i}$. 

\begin{figure}[ht]
    \centering
    \begin{minipage}{0.49\textwidth}
        \centering
        \includegraphics[width=1\textwidth]{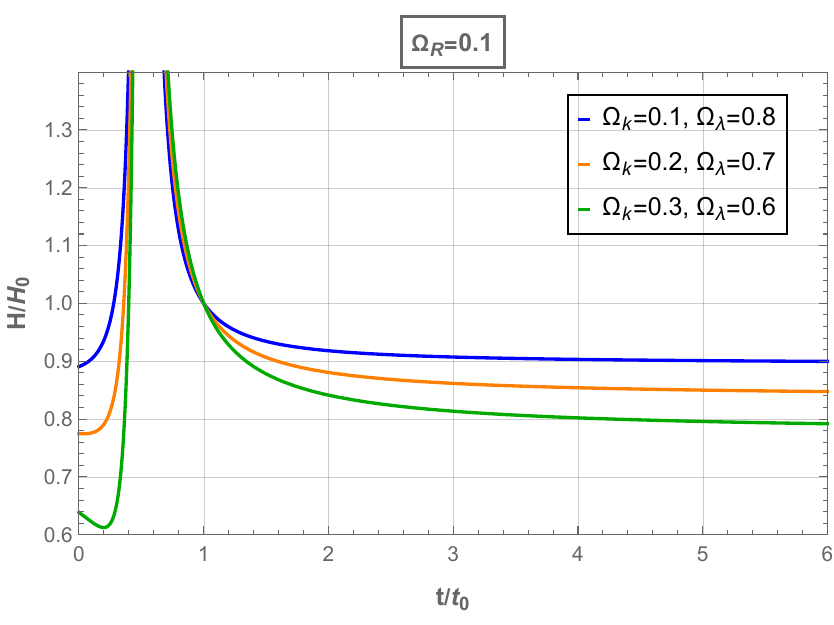}
        \caption{\small{Hubble parameter as a function of time for different values of $\{\Omega_{k},\Omega_{\lambda} \}$}.}
        \label{fig:1}
    \end{minipage}
    \hfill
    \begin{minipage}{0.49\textwidth}
        \centering
        \includegraphics[width=1\textwidth]{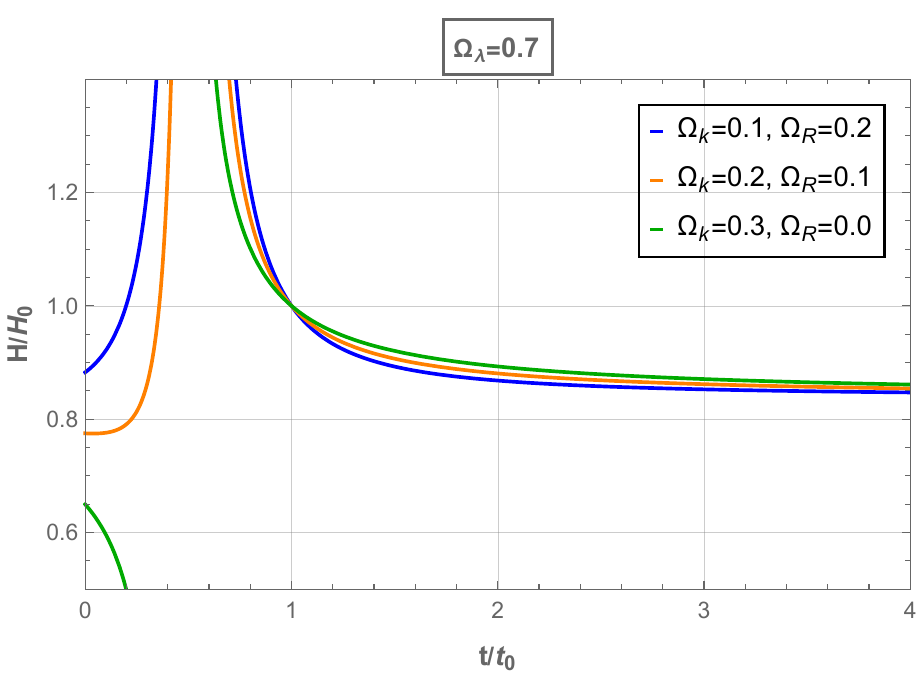}
        \caption{\small{Hubble parameter as a function of time for different values of $\{\Omega_{k},\Omega_{R} \}$}.}
        \label{fig:2}
    \end{minipage}
\end{figure}
Before proceeding towards the study of gravitational waves in the novel geometric condensate background, we need to address an important issue. Recall that the {\it{quadratic gravity model \cite{alv} reduced into a decoupled system of the graviton and a higher derivative scalar when expanded around a constant curvature (dS) background and that the latter yielded the condensate.}}  However, now we are employing this condensate as a background of FLRW scenario that has a non-constant curvature. One might question the validity of our approach. Below we provide a justification of our scheme. 

The Ricci scalar for an FLRW background ($k=0$) in terms of the scale factor is given by, 
\begin{eqnarray}
    R &=& 6 \bigg(\frac{\ddot{a}}{a} + \frac{\dot{a}^{2}}{a^{2}} \bigg) 
\end{eqnarray} 
Let us use the explicit form of the  scale factor, given in Appendix C, Eq. (\ref{scale_factor}), to compute the Ricci scalar $R$. $H_0(t-t_0)$ being a small dimensionless parameter,  we consider $O(H_0^2(t-t_0)^2)$ form of the scale factor, 
\begin{eqnarray}\label{approx_scale}
    a(t) &\approx& a(t_{0}) \sqrt{1 + 2H_{0}(t-t_{0}) + 2H^{2}_{0} \big(\Omega_{\lambda} + \Omega_{k}/2 \big) (t-t_{0})^{2}}.
\end{eqnarray}
The Ricci scalar of the background FLRW metric turns out to be 
\begin{eqnarray}
       R &=& 24 H^{2}_{0} \big(\Omega_{\lambda} + \Omega_{k}/2 \big) + \, O(H_0^3(t-t_0)^3).
\end{eqnarray}
The above exercise clearly demonstrates that the FLRW model in our lowest non-trivial approximation does have a constant curvature and hence the GC (generated in the quadratic gravity with  a constant background curvature) can be applicable here as well. Furthermore we find a positive $R$ that is consistent with the de Sitter spacetime considered in \cite{alv}, in this context.

\section{Solution of gravitational wave equation}

Let us return to the time dependent portion of GW equation
\begin{equation}
    \bigg(\frac{d^{2}v}{dt^{2}} \bigg) + \frac{3\dot{a}}{a} \bigg(\frac{dv}{dt} \bigg) - \bigg(\frac{2\ddot{a}}{a} + \frac{6\dot{a}^{2}}{a^{2}} - \frac{\Gamma}{a^{2}} \bigg) v = 0\label{tt1} 
\end{equation}
Till now, we did not use any particular functional form for the scale factor $a(t)$. Let us use the explicit form $a(t) = \sqrt{B+\gamma_0t}$, derived in the Appendix-C. If we put the previous solution of the scale factor in the time dependent part of gravitational wave equation (\ref{tt1}), we get  
\begin{eqnarray}\label{GW equation}
    &&  \bigg[ \bigg(\frac{d^{2}v}{dt^{2}} \bigg) + \hspace{0.1cm} 3 \sqrt{\frac{R_{0}c^{2}}{12} } \bigg\{\frac{1}{16} + \frac{\big(\alpha R_{0} \big)^{3/4}}{\big(B+\gamma_{0}t \big)} + \frac{\alpha R_{0}}{8 \big(B+\gamma_{0}t \big)^{2}} \bigg\}^{\boldsymbol{1/2}} \bigg(\frac{dv}{dt} \bigg) \nonumber \\ 
    && - \hspace{0.1cm} \frac{2R_{0}c^{2}}{4} \bigg\{\frac{1}{12} + \frac{\big(\alpha R_{0} \big)^{3/4} - \big(4\Gamma/R_{0}c^{2} \big)}{\big(B+\gamma_{0}t \big)} + \frac{\alpha R_{0}}{12 \big(B+\gamma_{0}t \big)^{2}} \bigg\} \hspace{0.1cm} v \bigg] = 0 .
    \label{dif}
\end{eqnarray}
The resulting equation in terms of cosmological density parameters $\Omega_{i}$ can be written as,
\begin{eqnarray}\label{Gravitational_wave}
    && \Bigg[\bigg(\frac{d^{2}v}{dt^{2}} \bigg) + \hspace{0.1cm} 3H_{0} \bigg\{\Omega_{\lambda} + \frac{\Omega_{k}}{1 + 2H_{0}(t-t_{0})} + \frac{\Omega_{R}}{\big(1 + 2H_{0}(t-t_{0}) \big)^{2}} \bigg\}^{\boldsymbol{1/2}} \bigg(\frac{dv}{dt} \bigg) \nonumber \\ 
    && \hspace{0.3cm} - \hspace{0.1cm} 2H^{2}_{0} \hspace{0.1cm} \bigg\{4\Omega_{\lambda} + \frac{3\Omega_{k} - \big(\Gamma/a^{2}_{0}H^{2}_{0} \big)}{1 + 2H_{0}(t-t_{0})} + \frac{2\Omega_{R}}{\big(1 + 2H_{0}(t-t_{0}) \big)^{2}} \bigg\} \hspace{0.1cm} v \Bigg] = 0 .
    \label{t12}
\end{eqnarray}
Our next task is to find the solution of (\ref{t12}) and investigate the effect of our GC background on the propagation of GW. In order to find $v(t)$, instead of using (\ref{t12}), it will be more convenient to return to the parent form (\ref{tt1}) written in terms of $a,\dot a$. This equation will be solved using a technique, elaborated in \cite{tech} (see Appendix B for details). The essential idea is that a generic equation of the form
\begin{equation}
\ddot{v} + \gamma(t) \hspace{0.07cm} \dot{v} + k(t) v = 0 
    \label{t1}
\end{equation}
can be expressed as 
\begin{equation}
\ddot{z} + \bigg(k(t) - \frac{\dot{\gamma}(t)}{2} - \frac{\gamma^{2}(t)}{4} \bigg) z = 0.
    \label{t2}
\end{equation}
where an explicit relation exists between the functions $v(t)$ and  $z(t)$ \cite{tech} (see Appendix B). In general, (\ref{t2}) represents a parametric oscillator with time dependent frequency, is more conducive to analytical solution. Comparing with (\ref{tt1}), the identification goes as  
\begin{eqnarray}
    \gamma(t) &=& \frac{3\dot{a}}{a} \nonumber \\ 
    k(t) &=& - \bigg(\frac{2\ddot{a}}{a} + \frac{6\dot{a}^{2}}{a^{2}} - \frac{\Gamma}{a^{2}} \bigg) .
    \label{tt2}
\end{eqnarray}
This leads to the   time dependent frequency $\omega(t)$
\begin{eqnarray}
    \omega^{2}(t) &=& \bigg[k(t) - \frac{\dot{\gamma}(t)}{2} - \frac{\gamma^{2}(t)}{4} \bigg] \nonumber \\ 
    &=& \boldsymbol{-} \frac{H^{2}_{0}}{4} \bigg[41 \hspace{0.07cm} \Omega_{\lambda} + \frac{\big(27 \hspace{0.07cm} \Omega_{k} - 4 \hspace{0.07cm} \Gamma/a^{2}_{0}H^{2}_{0} \big)}{1 + 2H_{0}(t-t_{0})} + \frac{13 \hspace{0.07cm} \Omega_{R}}{\big(1 + 2H_{0}(t-t_{0}) \big)^{2}} \bigg] .
    \label{fr}
\end{eqnarray}
where the explicit form of $a(t)$ is reinserted in (\ref{tt2}). Hence the differential equation to be solved, is 
\begin{equation}
    \ddot{z} - \frac{H^{2}_{0}}{4} \bigg[41 \hspace{0.07cm} \Omega_{\lambda} + \frac{\big(27 \hspace{0.07cm} \Omega_{k} - 4 \hspace{0.07cm} \Gamma/a^{2}_{0}H^{2}_{0} \big)}{1 + 2H_{0}(t-t_{0})} + \frac{13 \hspace{0.07cm} \Omega_{R}}{\big(1 + 2H_{0}(t-t_{0}) \big)^{2}} \bigg] z = 0 .
\end{equation}
Fortunately this is the   well known  Whittaker equation {\footnote{ Whittaker equation appears in diverse physical phenomena, (especially with problems enjoying spherical symmetry), such as radial wave equation in  Quantum Mechanics, Atomic and Molecular physics, central potential problems in General Relativity, wave propagation and stellar models of Astrophysics, to name a few.}} whose solutions can be written as a linear combination of the first and second kind of Whittaker functions $M_{A,B}(x)$ and $W_{A,B}(x)$ respectively. 
\begin{eqnarray}
    z(t) &=&  c_{1} \hspace{0.1cm} M_{\{A,B\}} \left(\frac{\sqrt{41\Omega_{\lambda}}}{2} \big(1 + 2H_{0}(t-t_{0}) \big) \right) + \hspace{0.1cm} c_{2} \hspace{0.1cm} W_{\{A,B\}} \left(\frac{\sqrt{41\Omega_{\lambda}}}{2} \big(1 + 2H_{0}(t-t_{0}) \big) \right)  . \hspace{1.4cm} 
\end{eqnarray}
Here $c_{1}$ and $c_{2}$ are the two integration constants of the second order differential equation. The indices of the Whittaker function in terms of the cosmological density parameters $\Omega_{i}$ are written as
\begin{equation}
    A = \bigg(\frac{4 \hspace{0.06cm} \Gamma - 27a^{2}_{0}H^{2}_{0} \hspace{0.06cm} \Omega_k}{8 a^{2}_{0} H^{2}_{0} \sqrt{41\Omega_{\lambda}}} \bigg) \hspace{0.6cm} ; \hspace{0.3cm} ~~B = -\frac{\sqrt{4 + 13\Omega_R}}{4} 
    \label{a1}
\end{equation}
The   Whittaker function are related to Hypergeometric $F_{1}(B,A,t)$ and Confluent Hypergeometric function $U_{1}(B,A,t)$ in the following way.   
\begin{eqnarray}
    M_{\{A,B \}}(t) &=& \exp{-t/2} \hspace{0.06cm} (t)^{B+1/2} \hspace{0.06cm} F_{1}\big(B-A+1/2 \hspace{0.06cm} ; 1+2B \hspace{0.06cm} ; t \big) \nonumber \\ 
    W_{\{A,B \}}(t) &=& \exp{-t/2} \hspace{0.06cm} (t)^{B+1/2} \hspace{0.06cm} U_{1}\big(B-A+1/2 \hspace{0.06cm} ; 1+2B \hspace{0.06cm} ; t \big) .
\end{eqnarray}
In the present problem, following Appendix B \cite{tech}, $v(t)= \frac{z(t)}{\sqrt{\Sigma(t)}} $ with $\Sigma(t) $ given by
\begin{eqnarray}
    \frac{1}{\Sigma} \hspace{0.1cm} \frac{d\Sigma}{dt} &=& \frac{3}{a} \hspace{0.1cm} \frac{da}{dt} \hspace{0.4cm} \rightarrow \hspace{0.4cm} 
    \Sigma(a) = \Sigma_{0} \bigg(\frac{a}{a_{0}} \bigg)^{3} .
\end{eqnarray}
It is now straightforward to derive the cherished form of  $v(t)$, time-dependent part of GW propagating on GC background,
\begin{eqnarray}
        v(t) \hspace{-0.24cm}~~ =~~ \hspace{-0.26cm} \big\{1+2H_{0}(t-t_{0}) \big\}^{-3/4} \bigg[ && \hspace{-0.6cm} c_{1} \hspace{0.1cm} M_{\{A,B\}} \left(\frac{\sqrt{41\Omega_{\lambda}}}{2} \big(1 + 2H_{0}(t-t_{0}) \big) \right) \nonumber\\
        + && \hspace{-0.6cm} c_{2} \hspace{0.1cm} W_{\{A,B\}} \left(\frac{\sqrt{41\Omega_{\lambda}}}{2} \big(1 + 2H_{0}(t-t_{0}) \big) \right) \bigg]. \hspace{0.7cm} 
\end{eqnarray}
This constitutes our principal result. The implications are discussed in the next section. In order to plot the GW amplitude as a function of time, we have to put the current value of Hubble constant $H_{0}$ and the age of the universe $t_{0}$ in the above expression. These values are: $H_{0} = 73.8 $ km/sec/Mpc and $t_{0} = 13.8 $ Billion years.{\footnote{The usual conversion rule of the astronomical units are the following: 1 Mpc $=$ $(3.26 \cross 10^{6})$ light years $=$ $(3.08 \cross 10^{19})$ km \hspace{0.06 cm} and \hspace{0.06 cm} 13.8 Billion years $=$ $(4.35 \cross 10^{17})$ sec.}} During plotting of the GW profile, we use the following values of the integration constants respectively: $c_{1}=0$ and $c_{2}=1.6$.

\begin{figure}[ht]
   \centering
   \begin{minipage}{0.49\textwidth}
       \centering
       \includegraphics[width=1\textwidth]{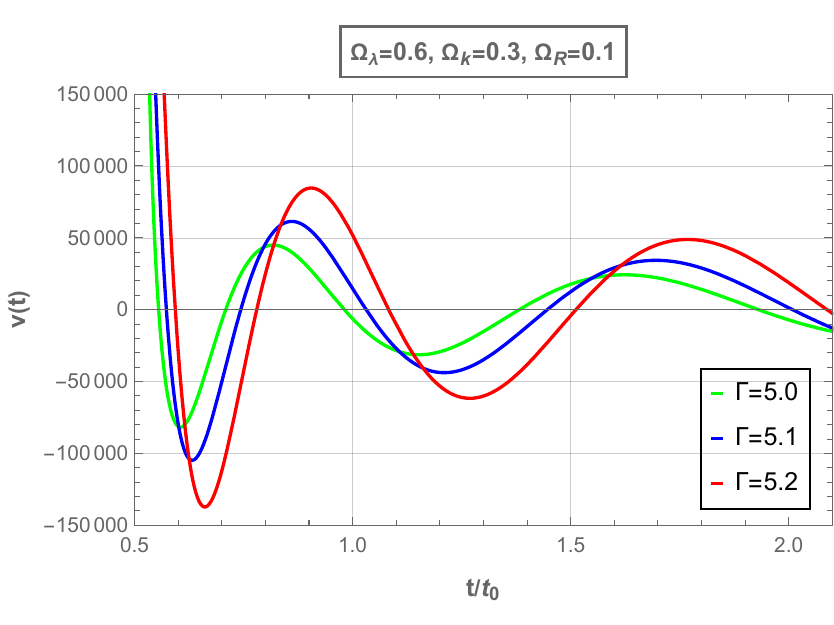}
       \caption{\small{Gravitational waves $v(t)$ as a function of time for different values of separation constant $\Gamma$.}}
       \label{fig:3}
   \end{minipage}
   \hfill
   \begin{minipage}{0.49\textwidth}
       \centering
       \includegraphics[width=1\textwidth]{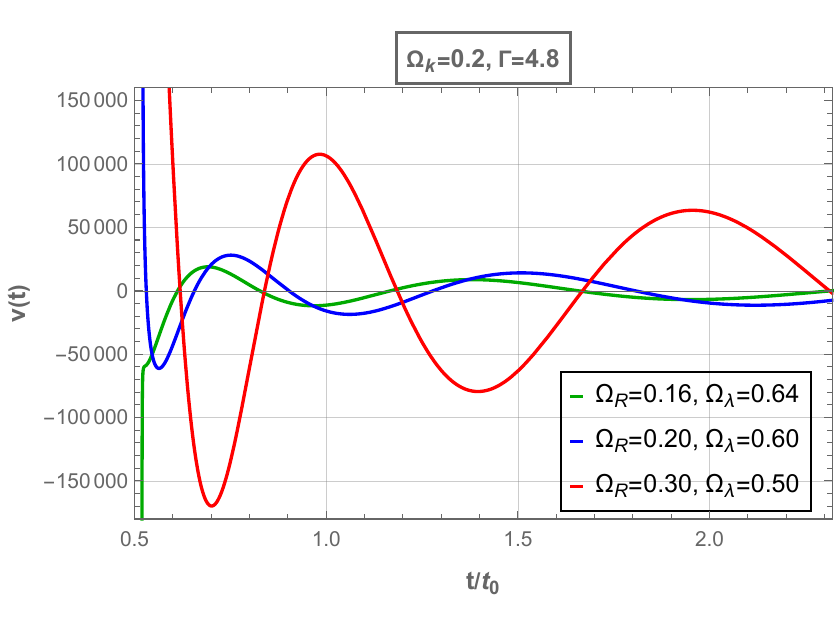}
       \caption{\small{Gravitational waves $v(t)$ as a function of time for different values of $\{\Omega_{R},\Omega_{\lambda} \}$.}}
       \label{fig:4}
   \end{minipage}
\end{figure}

\begin{figure}[ht]
    \centering
    \begin{minipage}{0.49\textwidth}
        \centering
        \includegraphics[width=1\textwidth]{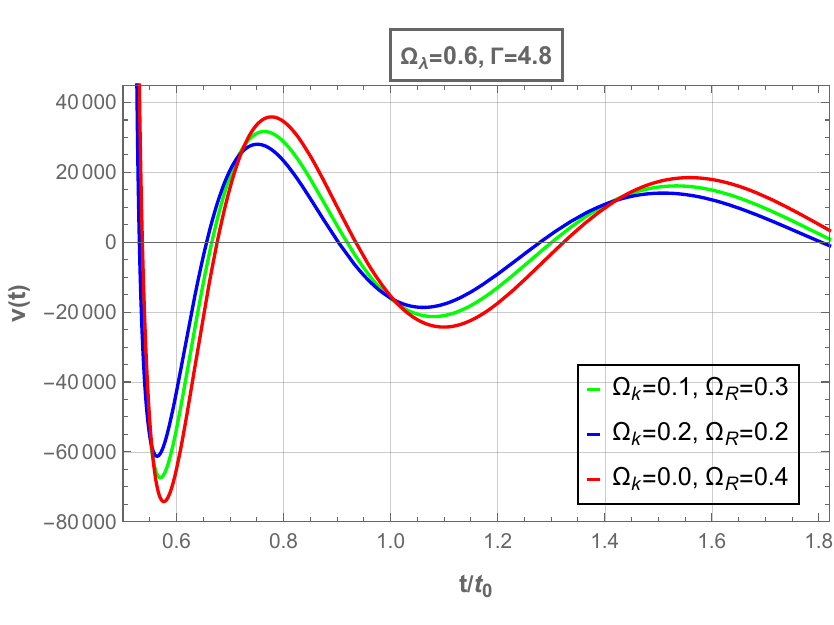}
        \caption{\small{Gravitational waves $v(t)$ as a function of time for different values of $\{\Omega_{k},\Omega_{R} \}$.}}
        \label{fig:5}
    \end{minipage}
    \hfill
    \begin{minipage}{0.49\textwidth}
        \centering
        \includegraphics[width=1\textwidth]{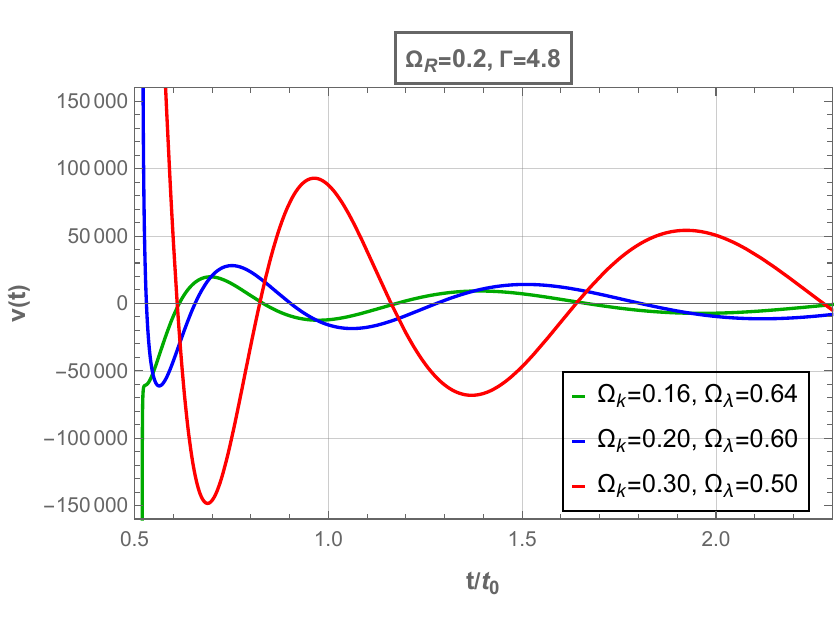}
        \caption{\small{Gravitational waves $v(t)$ as a function of time for different values of $\{\Omega_{k},\Omega_{\lambda} \}$.}}
        \label{fig:6}
    \end{minipage}
\end{figure}

\subsection{Dispersion relation}

Let us recall the GW equation propagating through FLRW background in presence of the geometrical condensate (GC) defined earlier. 
\begin{equation}
    \bigg(\frac{\partial^{2}h}{\partial t^{2}} \bigg) + \frac{3\dot{a}}{a} \bigg(\frac{\partial h}{\partial t} \bigg) - \frac{1}{a^{2}} \bigg(\frac{\partial^{2}h}{\partial x^{2}} \bigg) - \bigg(\frac{2\Ddot{a}}{a} + \frac{6\dot{a}^{2}}{a^{2}} \bigg) h = 0 
\end{equation}\label{GW_Eq}

The GW solution has two separable portions like $h(t,x) = v(t)u(x)$. To get an intuitive understanding of the oscillatory structure of $v(t)$, we note that limiting form of $W_{\{A,B \}}(t)$  can be approximated in terms of the usual trigonometric functions as given below.  
\begin{eqnarray}
        W_{\{A,B\}}(\Delta t) &\approx & - \bigg(\frac{4\Delta t}{A} \bigg)^{1/4} e^{-A+A\ln{(A)}} \hspace{0.1cm} \sin{\bigg\{2\sqrt{A\Delta t} - \big(4A+1 \big)\frac{\pi}{4} \bigg\}} .
    \label{b1}
\end{eqnarray}
This leads to the limiting form of $v(t)$, time dependent part of GW as
\begin{eqnarray}
    v(t) &\approx& - c_{2} \hspace{0.1cm} e^{-A+A\ln{(A)}} \bigg(\frac{2 \sqrt{41\Omega_{\lambda}}}{A^{1/3}} \bigg)^{3/4} \frac{1}{\sqrt{4\Delta t}} \hspace{0.1cm} \sin{\bigg\{2\sqrt{A\Delta t} - \big(4A+1 \big)\frac{\pi}{4} \bigg\}} . \hspace{1.0cm} 
    \label{Approx_GW}
\end{eqnarray}
Here $A,B$ are defined as in (\ref{a1}) and $\Delta t$ is given by 
\begin{equation}
    \Delta t \hspace{0.2cm} = \hspace{0.2cm} \frac{\sqrt{41\Omega_{\lambda}}}{2} \big\{1 + 2H_{0}(t-t_{0}) \big\} .
\end{equation}
The condition under which the approximation are valid is
\begin{eqnarray}
       \big(27 \hspace{0.06cm} \Omega_{k} - 164 \hspace{0.06cm} \Omega_{\lambda} \big) &\gg& \bigg(\frac{4\Gamma}{a^{2}_{0}H^{2}_{0}} \bigg) .
\end{eqnarray}
Due to the time dependent and non-trivial nature of the GC background, GWs are dispersive in nature with an involved dispersion relation. The solution of the GW profile will become, 
\begin{eqnarray}\label{GW_solution}
    h(t,x) &=& v(t) \hspace{0.05cm} u(x) \nonumber \\ 
    &\approx& c_{3} c_{2} \hspace{0.1cm} e^{-A+A\ln{(A)}} \bigg(\frac{2 \sqrt{41\Omega_{\lambda}}}{A^{1/3}} \bigg)^{3/4} \hspace{0.1cm} \frac{e^{-i\sqrt{\Gamma}x}}{\sqrt{4\Delta t}} \hspace{0.1cm} \sin{\bigg\{2\sqrt{A\Delta t} - \big(4A+1 \big)\frac{\pi}{4} \bigg\}} \nonumber \\ 
    &\approx& c_{3} c_{2} \hspace{0.1cm} e^{-A+A\ln{(A)}} \bigg(\frac{2 \sqrt{41\Omega_{\lambda}}}{A^{1/3}} \bigg)^{3/4} \hspace{0.1cm} \frac{e^{-i\sqrt{\Gamma}x}}{\sqrt{4\Delta t}} \hspace{0.1cm} \sin{\bigg\{\omega(t) \Delta t - (4A+1)\frac{\pi}{4} \bigg\}}
\end{eqnarray}
Here $c_{2}$ and $c_{3}$ are  integration constants which may be specified by using some appropriate boundary conditions. Expressing  the oscillatory part of the above solution in the form $``\sin(\omega \Delta t)" $, one can  extract an effective frequency $\omega(t)$ and an effective wave vector $k$ for GW as given below. 
\begin{eqnarray}
    \omega(t) &=& \frac{1}{a_{0}} \sqrt{\frac{4\Gamma - 27 
    \hspace{0.05cm} \Omega_{k}a^{2}_{0}H^{2}_{0}}{1 + 2H_{0}(t-t_{0})}} \hspace{0.6cm} ; \hspace{0.6cm}  k \hspace{0.2cm} = \hspace{0.2cm} \sqrt{\Gamma} .
\end{eqnarray}
Note that the separation constant $\Gamma$, that appeared   (\ref{uv}) in the differential equation obeyed by $u(x)$, the spatial part of GW profile $h(t,x)$,  is simply the squared effective wave vector $k^{2}$, a possible observable in GW phenomena. 

Substituting the above solution of GW, given by (\ref{GW_solution}), in the partial differential equation (\ref{g2}), one gets the following non trivial dispersion relation. 
\begin{eqnarray}\label{Dispersion_1}
    \frac{\omega^{2}}{4} - \frac{k^{2}}{a(t)^{2}} &=& \bigg[\frac{3\omega H_{0}}{2} \cot{\bigg\{\frac{\omega}{2H_{0}} \frac{a(t)^{2}}{a_{0}^{2}} - (4A+1)\frac{\pi}{4} \bigg\}} - \frac{3H^{2}_{0}}{\left(a(t)/a_{0} \right)^{2}} \bigg] \bigg[\frac{H(t)}{H_{0}} - \frac{a_{0}^{2}}{a(t)^{2}} \bigg] \nonumber \\ 
    && - \hspace{0.1cm} 2H^{2}_{0} \bigg[4\Omega_{\lambda} + \frac{3\Omega_{k}}{\left(a(t)/a_{0} \right)^{2}} + \frac{2\Omega_{R}}{\left(a(t)/a_{0} \right)^{4}} \bigg] . 
\end{eqnarray}
The scale factor $a(t)$, present in the above dispersion relation of gravitational waves (\ref{Dispersion_1}), can be written as follows, as given in the Appendix \ref{approx_scale_factor}, 
\begin{equation}
    a(t) \approx a_{0} \sqrt{1 + 2H_{0}(t-t_{0})} \hspace{0.06cm} . 
\end{equation}
The above dispersion relation can be further simplified through the below  transformation of frequency, $\overline{\omega}=\omega(t) a(t)/2 \hspace{0.07cm} = \hspace{0.07cm} \sqrt{4\Gamma - 27 \hspace{0.05cm} \Omega_{k}a^{2}_{0}H^{2}_{0}}/2$. 
\begin{eqnarray}\label{Dispersion_2}
    \overline{\omega}^{2} - k^{2} &=& \bigg[3 \hspace{0.06cm} \overline{\omega}H_{0} a(t) \cot{\bigg\{\frac{\overline{\omega}}{a_{0}H_{0}} \frac{a(t)}{a_{0}} - (4A+1)\frac{\pi}{4} \bigg\}} - 3a^{2}_{0}H^{2}_{0} \bigg] \bigg[\frac{H(t)}{H_{0}} - \frac{a_{0}^{2}}{a(t)^{2}} \bigg] \nonumber \\ 
    && - \hspace{0.1cm} 2H^{2}_{0}a(t)^{2} \bigg[4\Omega_{\lambda} + \frac{3\Omega_{k}}{\left(a(t)/a_{0} \right)^{2}} + \frac{2\Omega_{R}}{\left(a(t)/a_{0} \right)^{4}} \bigg] . \hspace{0.7cm}  
\end{eqnarray}
Here $H(t)$ be the Hubble parameter of our model given earlier (\ref{Hubble}). The involved nature of the GW dispersion relation arises principally due to its propagation in the Geometric Condensate background. Generalized forms of GW dispersion have appeared in extended gravity models (see for example \cite{ser1}).

\section{Discussion}

Out of the two independent solutions of Whittaker equation, $M_{\{A,B \}}(t)$ and $W_{\{A,B \}}(t)$, the former has a diverging nature for large values of its arguments (time in our case) whereas the latter is well-behaved and  converging throughout the full range of time. Therefore, from physical point of view, we drop  $M_{\{A,B \}}(t)$ and consider only $W_{\{A,B \}}(t)$ as the solution of GW equation. Furthermore, the solution of time dependent portion of GW $v(t)$ would yield a real quantity only when the following condition is satisfied,  
\begin{eqnarray}\label{constraint_1}
    && 1 + 2H_{0}(t-t_{0}) \hspace{0.2cm} > \hspace{0.2cm} 0 \nonumber \\ 
    && \hspace{0.1cm} t/t_{0} \hspace{0.2cm} > \hspace{0.2cm} \bigg( \frac{2H_{0}t_{0}-1}{2H_{0}t_{0}} \bigg) .  
\end{eqnarray} 
Using the current observational values of Hubble constant $H_{0}$ and the universe's age $t_{0}$, the above limit comes out to be $t/t_{0} > 0.52$. The time-scale over which the above condition is satisfied, yields the real solution of the GW equation. Thus, as also verified from the figures,  earlier times such that $t/t_{0} < 0.52$ are not accessible in the present analysis. However, as we noted earlier, the numerical values are mentioned just as a benchmark and may not be realistic enough. 

Recall that the  approximation used earlier in (\ref{yy},\ref{y}), (for details see Appendix C),  $2\sqrt{R_0c^2/192}(t-t_0)<1$ can be written in the form 
\begin{eqnarray}\label{constraint_2}
    && 2H_{0} \sqrt{\Omega_{\lambda}} (t-t_{0}) \hspace{0.2cm} < \hspace{0.2cm} 1 \nonumber \\ 
    && t/t_{0} \hspace{0.02cm} < \hspace{0.2cm} \left(\frac{1 + 2H_{0}t_{0}\sqrt{\Omega_{\lambda}}}{2H_{0}t_{0}\sqrt{\Omega_{\lambda}}} \right) . 
\end{eqnarray} 
Unlike the previous condition  (\ref{constraint_1}), the above condition  (\ref{constraint_2}) depends on the density parameter $\Omega_{\lambda}$ that corresponds to the effective cosmological constant. If one uses realistic and present day values of  $\Omega_{\lambda}=0.7$, the Hubble parameter $H_0$ and the age of the universe $t_0$, then the upper limit of the time-scale is found to be $t/t_{0} < 1.57$.  Using both of these conditions, the range of validity of our  GC  cosmological model is $0.52 < t/t_{0} < 1.57$.  Notice that, the upper bound of the time-scale over which our GC kind of model is valid, explicitly depends on the density parameter $\Omega_{\lambda}$. 

Through figures \hyperref[fig:3]{$(3)$}-\hyperref[fig:6]{$(6)$}, time dependent part of the GW propagating through the GC background is plotted against time, scaled by $t_0$. In the figure \hyperref[fig:3]{$(3)$}, GW profile  $v(t)$ is plotted for three distinct values of the separation constant ${\Gamma}$, keeping  the density parameters $\Omega_{i}$'s fixed. This shows that the arbitrariness in separation constant $\Gamma$ does not alter the GW profile in a qualitative way but affects quantitatively and $\Gamma$ can be fixed by comparing with observations. In figures \hyperref[fig:4]{$(4)$}-\hyperref[fig:6]{$(6)$}, ${\Gamma}=4.8$ is kept unchanged throughout. In each of these figures \hyperref[fig:4]{$(4)$}, \hyperref[fig:5]{$(5)$} and \hyperref[fig:6]{$(6)$} respectively, the density parameters $\{\Omega_k,\Omega_\lambda,\Omega_R \}$ are  varied accordingly. Indeed the $\Omega_i$'s are varied subject to the constraint (\ref{om}). The plots show that, as also expected from (\ref{fr}), the frequency of the GW is most sensitive to the variations of $\Omega_{\lambda}$ less so for $\Omega_{k}$ and $\Omega_{R}$.

To summarize, the present work deals with the propagation of Gravitational Waves through a condensate background. It is important to note that the condensate is not static; it oscillates in time and is originated from the Starobinsky-like $R^2$-gravity model, as a higher derivative scalar excitation. This is quite akin to a Time Crystal scenario. Presence of this condensate was conjectured earlier in \cite{ssg} and cosmology in this background was studied in \cite{us}. What separates our model from other existing condensate models in cosmology is that in all the latter cases the condensate is formed by matter, introduced from outside whereas no such matter is considered in our model. The condensate is generated purely from gravitational degrees of freedom, in the form of a higher derivative scalar, and hence is termed as a geometric condensate. 

The central difference between our GC and other condensate models of cosmology like \textit{BEC, Group Field Theory condensate}, is that in the later, the matter contribution is added from outside as an external degree of freedom. But in the present context, the \textit{Geometric Condensate} is constructed completely from the scalar degree of freedom of the Starobinsky-like $\left(R + \alpha R^{2} \right)$-action upon metric perturbations. The primary novelty of this GC is that it is originated from the geometric degrees of freedom of a higher derivative gravitational theory (that is free from any kind of Ghost fields), not a matter content, introduced from outside. 

We have shown that this geometric condensate is capable of producing spatial curvature and radiation like contributions in FLRW cosmology. It will be interesting to see if more careful analysis can lead to matter like contributions as well. On the other hand, one can include matter contributions to consider more realistic cosmological scenario. 

In the cosmological scenario, the non-trivial time dependency of GC is primarily responsible for the generation of the effective spatial curvature $\Omega_{k}$ and effective radiation $\Omega_{R}$ like behaviour in the expansion history of the universe. It actually ensures the possible existence of this kind of GC as a background matter content of the FLRW universe. Remember that, it is basically independent of the existence of any barotropic fluid. Secondly, we investigated the propagation of Gravitational Waves through FLRW background in presence of the above mentioned GC. It significantly changes the Dispersion relation of GW, given by Eq. (\ref{Dispersion_2}), through the presence of the effective spatial curvature $\Omega_{k}$ and radiation $\Omega_{R}$ like terms. Therefore, in principle, the signature of GC background, present in the right hand side of the dispersion Relation Eq. (\ref{Dispersion_2}), could be extracted through the observational detection of GW. These two are the central results of this current manuscript which would leave its imprint in the observational parameters of cosmology, like density parameters $\Omega_{i}$'s and GW events, like dispersion relation $\left(\omega-k \right)$.  

Lastly our idea of Geometric Condensate is "cleaner" and more down to earth in comparison with the  diverse condensate models in cosmological context. The latter  are afflicted with a common weakness  that the generation of the condensates themselves rely on  some exotic mechanisms or unsubstantiated matter content. The present GC scores above the others precisely in this context: we only introduce the quadratic ($R^2$) Starobinski extension on top of  Einstein gravity and that is sufficient to induce the GC in purely conventional physics principles.    \\

\appendix
\noindent\textbf{\Large{Appendices}}

\section{Gravitational wave equation} 

Let us begin the study of gravitational waves through the introduction of a linear perturbation on an arbitrary curved spacetime. Generally these perturbations are introduced for achieving results that are very close to the real physical world. The entire perturbed metric can be written as
\begin{equation}
    g_{\mu\nu} = \overline{g}_{\mu\nu} + \epsilon h_{\mu\nu}
\end{equation}
Here $\epsilon$ is a dimensionless small parameter attached with the linear perturbation of the metric. On the other hand, the contravariant component of metric tensor can be found as: $g^{\mu\nu}=\big(\overline{g}^{\mu\nu} - \epsilon h^{\mu\nu} \big)$. Both of these components of the metric tensor are required for doing differential geometry on a curved background. Now our aim is to find the exact form of Einstein field equation in the background and perturbation level both. Using these above definitions, the form of Christoffel connection is found to be the following. 
\begin{eqnarray}
    \Gamma^{\alpha}_{\beta\mu} &=& \overline{\Gamma}^{\alpha}_{\beta\mu} + \delta\Gamma^{\alpha}_{\beta\mu}
\end{eqnarray}
Here $\delta\Gamma^{\alpha}_{\beta\mu}$ arises purely from perturbed metric of our entire spacetime. It's mathematical form is: $\delta\Gamma^{\alpha}_{\beta\mu} = \frac{1}{2}\overline{g}^{\alpha\nu}\big(\nabla_{\mu}h_{\beta\nu} + \nabla_{\beta}h_{\mu\nu} - \nabla_{\nu}h_{\beta\mu} \big)$. Once we know the Christoffel connection, one can easily define the covariant derivative of any tensorial quantity over this curved manifold. In order to find the curvature, one has to concentrate on the Riemann tensor and its other counterparts. The quantity that appears in the expression of Einstein field equations, are Ricci tensor and Ricci scalar respectively. After a rigorous calculation, one finds the following. 
\begin{equation}
    R_{\alpha\beta} = \overline{R}_{\alpha\beta} + \delta R_{\alpha\beta}
\end{equation}
Here $\overline{R}_{\alpha\beta}$ is the background contribution of the metric tensor on the once contracted curvature tensor. And $\delta R_{\alpha\beta}$ is constructed purely from the perturbed metric as follows: $\delta R_{\alpha\beta} = \big(\nabla_{\mu}\nabla_{\alpha}h^{\mu}_{\beta} + \nabla_{\mu}\nabla_{\beta}h^{\mu}_{\alpha} - \Box h_{\alpha\beta} - \nabla_{\beta}\nabla_{\alpha}h \big)$. Using these above quantities, one can explicitly find the form of Einstein field equations both in background and perturbation level. It looks like the following. 
\begin{eqnarray}
    R_{\alpha\beta} &=& 8\pi G \big(T_{\alpha\beta} - \frac{1}{2}g_{\alpha\beta}T \big) \nonumber \\ 
    &=& 8\pi G \big[\big(\overline{T}_{\alpha\beta} - \frac{1}{2}g_{\alpha\beta}\overline{T} \big) + \big(\delta T_{\alpha\beta} - \frac{1}{2}\overline{g}_{\alpha\beta}\delta T \big) \big]
\end{eqnarray}
Generally it is an usual practice to take the perturbation on both sectors of gravitational field equation, geometry and matter. Here $\overline{T}_{\alpha\beta}$ and $\delta T_{\alpha\beta}$ are the background and perturbative stress-energy tensor of our matter content respectively. For simplicity, lets us ignore the perturbative part of the stress-energy tensor i.e. $\delta T_{\alpha\beta} \approx 0$ \cite{maggiore,nb}. In the absence of perturbed matter field, the linear order perturbed portion of Einstein field equation becomes
\begin{eqnarray}
    && \delta R_{\alpha\beta} = 8\pi G \big(\delta T_{\alpha\beta} - \frac{1}{2}\overline{g}_{\alpha\beta} \hspace{0.06cm} \delta T \big) \nonumber \\ 
    && \delta R_{\alpha\beta} = 0 \nonumber \\ 
    && \big(\nabla_{\mu}\nabla_{\alpha}h^{\mu}_{\beta} + \nabla_{\mu}\nabla_{\beta}h^{\mu}_{\alpha} - \Box h_{\alpha\beta} - \nabla_{\beta}\nabla_{\alpha}h \big) = 0 
\end{eqnarray}
For the purpose of simplification, it is better to rewrite this entire equation in terms of trace reversed tensor $\Tilde{h}_{\alpha\beta}$ whose definition is this: $\Tilde{h}_{\alpha\beta} = \big(h_{\alpha\beta} - \frac{1}{2} \overline{g}_{\alpha\beta
} h \big)$. Apart from this, in order to remove the gauge redundancy of our gravitational theory, we have to impose some constrains on the gravitational degrees of freedom $\Tilde{h}_{\alpha\beta}$. One of the most popular choice is the Lorenz gauge whose mathematical form is as follows: $\nabla_{\alpha}\Tilde{h}^{\alpha\beta}=0$. After using all these transformations and constrains on the perturbative field equations, one gets the following final form. 
\begin{equation}
    \big(\Box\Tilde{h}_{\alpha\beta} - 2\overline{R}_{\nu\beta\alpha\mu} \Tilde{h}^{\nu\mu} - \overline{R}_{\nu\alpha}\Tilde{h}^{\nu}_{\beta} - \overline{R}_{\nu\beta}\tilde{h}^{\nu}_{\alpha} \big) = 0 
\end{equation}
It is the final form of gravitational wave equation propagating on any curved background. The background Riemann tensor arises from the commutation relation of successive covariant derivatives already present in this above equation. The trace of the perturbed metric $h_{\alpha\beta}$ is also absent in the final wave equation. It happens because we are using traceless transverse gauge (TT gauge) in which the trace of the perturbed metric is usually taken to be zero i.e. $\tilde{h}=h=0$. That's why it is called traceless gauge. On the other hand, the name ``Transverse" is justified through the Lorenz gauge condition.

\section{Conversion of a damped oscillator into parametric oscillator} 

In order to solve the differential equation of a damped harmonic oscillator with time dependent coefficients, we are going to follow the technique of this reference \cite{tech}. Let us give a brief introduction of the technique. Our main goal is to find a canonical transformation through which the equation of motion of a damped harmonic oscillator has been converted to that of a harmonic oscillator with time dependent frequency. Depending on the functional form of this frequency, one can find the solution of the corresponding simplified equation of motion. \\ 

Let us start with the following Lagrangian consisting of two independent degrees of freedom $``x"$ and $``y"$ respectively. 
\begin{equation}
    \mathcal{L}(x,y,\dot{x},\dot{y}) = \big[\dot{x}\dot{y} + \frac{\gamma(t)}{2} (x\dot{y} - y\dot{x}) + \bigg(\frac{\dot{\gamma}(t)}{2} - k(t) \bigg) xy \big]
\end{equation}
Here $\gamma(t)$ and $k(t)$ are two arbitrary functions of time. If one finds the equation of motion corresponding to the above Lagrangian, then we find that it coincides with that of a damped harmonic oscillator for both degrees of freedom. Its mathematical form is somewhat like this. 
\begin{eqnarray}\label{Damped_oscillator}
    && \ddot{x} + \gamma(t) \hspace{0.07cm} \dot{x} + k(t)x = 0 \nonumber \\ 
    && \ddot{y} - \gamma(t) \hspace{0.07cm} \dot{y} + \{k(t) - \dot{\gamma}(t)\} y = 0 
\end{eqnarray}
Here the second differential equation for $y(t)$ is time reversed version of the first one. The reason for choosing such a Lagrangian is clear from the equation of motion of $x(t)$ which appears exactly in the context of gravitational waves propagating on a cosmological background. Before doing any kind of canonical transformation of the conjugate variables, it is desirable to find the functional form of the Hamiltonian. For our given Lagrangian, one can easily find out the Hamiltonian through its usual definition.  
\begin{eqnarray}
    H(x,p_{x},y,p_{y}) &=& \big[p_{x}\dot{x} + p_{y}\dot{y} - \mathcal{L}(x,y,\dot{x},\dot{y}) \big] \nonumber \\ 
    &=& \big[p_{x}p_{y} + \frac{\gamma(t)}{2} \big(p_{y}y - p_{x}x \big) + \bigg(k(t) - \frac{\dot{\gamma}(t)}{2} - \frac{\gamma^{2}(t)}{4} \bigg) xy \big]
\end{eqnarray}
The presence of the crossed terms between the phase space variables $\{x,p_{x},y,p_{y} \}$ in the expression of the above Hamiltonian, are mainly responsible for the existence of single derivative terms $(\dot{x},\dot{y})$ in the equation of motion. In order to convert the damped harmonic oscillator into a simple harmonic oscillator, we have to do such a canonical transformation which can remove the crossed terms of phase space variables from the expression of modified Hamiltonian. For this purpose, let us choose the following kind of generating function $F(x,y,p_{1},p_{2})$. 
\begin{equation}
    F(x,y,p_{1},p_{2}) = \bigg[\sqrt{\Sigma(t)} \hspace{0.1cm} xp_{1} + \frac{yp_{2}}{\sqrt{\Sigma(t)}} \bigg] 
\end{equation}
This generating function is responsible for creating the canonical transformation from one set of variables $\{x,p_{x},y,p_{y} \}$ to other $\{z_{1},p_{1},z_{2},p_{2} \}$. Up to now, $\Sigma(t)$ is an arbitrary function of time which can be evaluated later in terms of Lagrangian coefficients $\gamma(t)$. Our next job is find the relation between the new and old set of phase space variables. 
\begin{eqnarray}\label{Canonical_trans}
    z_{1} &=& \frac{\partial F}{\partial p_{1}} = \sqrt{\Sigma(t)} \hspace{0.1cm} x \hspace{0.5cm} ; \hspace{0.5cm} z_{2} \hspace{0.1cm} = \hspace{0.1cm} \frac{\partial F}{\partial p_{2}} = \frac{y}{\sqrt{\Sigma(t)}} \nonumber \\ 
    p_{x} &=& \frac{\partial F}{\partial x} = \sqrt{\Sigma(t)} \hspace{0.1cm} p_{1} \hspace{0.5cm} ; \hspace{0.5cm} p_{y} \hspace{0.1cm} = \hspace{0.1cm} \frac{\partial F}{\partial y} = \frac{p_{2}}{\sqrt{\Sigma(t)}} 
\end{eqnarray}
Our new Hamiltonian in terms of these new set of variables can be written as below. 
\begin{eqnarray}
    \mathcal{H}(z_{1},p_{1},z_{2},p_{2}) &=& \bigg[H(x,p_{x},y,p_{y}) + \frac{\partial F}{\partial t} \bigg] \nonumber \\ 
    &=& \bigg[p_{1}p_{2} + \bigg(k(t) - \frac{\dot{\gamma}(t)}{2} - \frac{\gamma^{2}(t)}{4} \bigg) z_{1}z_{2} + \bigg\{\frac{\gamma(t)}{2} - \frac{\partial \Sigma(t)/\partial t}{2\Sigma(t)} \bigg\} \big(z_{2}p_{2} - z_{1}p_{1} \big) \bigg] \hspace{0.8cm} 
\end{eqnarray}
In order to remove the crossed term from our new Hamiltonian, we have to make its coefficients vanish by imposing a constrain on the arbitrary function $\Gamma(t)$. The constrain is the following. 
\begin{equation}\label{Constrain}
    \frac{1}{\Sigma(t)} \frac{\partial \Sigma(t)}{\partial t} \hspace{0.1cm} = \hspace{0.1cm} \gamma(t)
\end{equation}
After using the above condition, we get a modified form of the new Hamiltonian without any crossed term. 
\begin{equation}
    \mathcal{H}(z_{1},p_{1},z_{2},p_{2}) = \bigg[p_{1}p_{2} + \bigg(k(t) - \frac{\dot{\gamma}(t)}{2} - \frac{\gamma^{2}(t)}{4} \bigg) z_{1}z_{2} \bigg]
\end{equation}
Now if one finds the equation of motion for these new set of conjugate variables $\{z_{1},z_{2} \}$, then we find its resemblance with that of a parametric oscillator. 
\begin{eqnarray}
    && \ddot{z_{1}} + \bigg(k(t) - \frac{\dot{\gamma}(t)}{2} - \frac{\gamma^{2}(t)}{4} \bigg) z_{1} = 0 \nonumber \\ 
    && \ddot{z_{2}} + \bigg(k(t) - \frac{\dot{\gamma}(t)}{2} - \frac{\gamma^{2}(t)}{4} \bigg) z_{2} = 0 
\end{eqnarray}
Here $z_{1}(t)$ corresponds to the time dependent, modified GW amplitude $z(t)$ written in the Eq. (\ref{t2}) of main text.

\section{Approximation of scale factor}\label{approx_scale_factor}  

Here we give a detailed derivation of the scale factor $a(t)$ in presence of an effective Cosmological constant, radiation and spatial curvature kind of matter content, used in the Eq. (\ref{approx_scale}) of the main text. For the small values of the hyperbolic arguments $2\sqrt{D}(t-t_{0})$, the cosine and sine hyperbolic functions are approximated as follows, $\cosh{\big(2\sqrt{D}(t-t_{0}) \big)} \approx \big\{1 + 2D(t-t_{0})^{2} \big\}$ \hspace{0.1cm} and \hspace{0.1cm} $\sinh{\big(2\sqrt{D}(t-t_{0}) \big)} \approx 2\sqrt{D}(t-t_{0})$. Using this approximation, the scale factor of our given cosmological model can be written as given below 
\begin{eqnarray}\label{scale_factor}
    a(t) &=& \bigg[m_{0} \hspace{0.1cm} \sinh{\bigg\{\sinh^{-1}{\frac{y_{0}}{m_{0}}} + 2\sqrt{D}(t-t_{0}) \bigg\}} - \frac{F}{2D} \bigg]^{1/2} \nonumber \\ 
    &=& \bigg[m_{0} \bigg\{\sinh{\bigg(\sinh^{-1}{\frac{y_{0}}{m_{0}}} \bigg)} \cosh{\big(2\sqrt{D}\delta t \big)} + \cosh{\bigg(\sinh^{-1}{\frac{y_{0}}{m_{0}}} \bigg)} \sinh{\big(2\sqrt{D}\delta t \big)} \bigg\} - \frac{F}{2D} \bigg]^{1/2} \nonumber \\ 
    &=& \bigg[y_{0} \hspace{0.1cm} \cosh{\big(2\sqrt{D}\delta t \big)} + \sqrt{y^{2}_{0}+m^{2}_{0}} \hspace{0.1cm} \sinh{\big(2\sqrt{D}\delta t \big)} - \frac{F}{2D} \bigg]^{1/2} \nonumber \\ 
    &\approx& \bigg[y_{0} \big\{1 + 2D \hspace{0.04cm} (t-t_{0})^{2} \big\} + \sqrt{y^{2}_{0}+m^{2}_{0}} \hspace{0.06cm} \big\{2\sqrt{D}(t-t_{0}) \big\} - \frac{F}{2D} \bigg]^{1/2} \nonumber \\ 
    &\approx& \bigg[\bigg(y_{0} - \frac{F}{2D} \bigg) + 2(t-t_{0}) \sqrt{D \big(y^{2}_{0}+m^{2}_{0} \big)} + 2y_{0}D \hspace{0.04cm} (t-t_{0})^{2} \bigg]^{1/2} \nonumber \\ 
%    &\approx& \sqrt{B + \gamma_{0}t} \nonumber \\ 
    &\approx& a(t_{0}) \sqrt{1 + 2H_{0} \big(t-t_{0} \big) + 2H^{2}_{0} \hspace{0.04cm} \big(\Omega_{\lambda}+\Omega_{k}/2 \big) \big(t-t_{0} \big)^{2}} . 
\end{eqnarray}
It is the approximated expression of the scale factor $a(t)$ in presence of our previously mentioned GC. The definition of the symbols $\{y_{0},m_{0},F,D \}$ used in the above derivation are given below. 
\begin{eqnarray}
    y_{0} &=& \big[a^{2}(t_{0}) + 8 \big(\alpha R_{0} \big)^{3/4} \big] \hspace{0.3cm} ; \hspace{0.3cm} m^{2}_{0} \hspace{0.1cm} = \hspace{0.1cm} \big[2 \alpha R_{0} - 64 \big(\alpha R_{0} \big)^{3/2} \big] \nonumber \\ 
    F &=& \frac{R_{0}c^{2}}{12} \big(\alpha R_{0} \big)^{3/4} \hspace{1.2cm} ; \hspace{0.3cm} D \hspace{0.1cm} = \hspace{0.1cm} \frac{R_{0}c^{2}}{192} \hspace{0.5cm} ; \hspace{0.5cm} \delta t = (t-t_{0}) .
\end{eqnarray}
Here $\alpha$ be the coupling constant, $R_{0}$ be the constant curvature associated with background metric and $a(t_{0})$ be the scale factor at initial time $t=t_{0}$.

%\vspace{1cm}
{\bf{Acknowledgement:}} We thank Professor Sergei Odintsov for informing us about related earlier works concerning GW in $f(R)$ models.

%\printbibliography

\end{document}